# Perfect optical absorption-enhanced magneto-optic Kerr effect microscopy

**Authors**: Dongha Kim[1,3], Young-Wan Oh[2,3], Jong-Uk Kim[2,3], Jonghwa Shin[2,3], Kab-Jin Kim[1], Byong-Guk Park[2,3], and Min-Kyo Seo[1,3*]

**Affiliation**:

[1]Department of Physics, KAIST, Daejeon 34141, Republic of Korea,

[2]Department of Materials Science and Engineering, KAIST, Daejeon 34141, Republic of Korea,

[3]Institute for the NanoCentury, KAIST, Daejeon 34141, Republic of Korea,

*Corresponding author: minkyo_seo@kaist.ac.kr

## Introduction

Magnetic and spintronic media have offered fundamental scientific subjects and technological applications[1]. Magneto-optic Kerr effect (MOKE) microscopy[2] provides the most accessible platform to study the dynamics of spins, magnetic quasi-particles, and domain walls[3,4,5,6]. However, in the research of nanoscale spin textures and state-of-the-art spintronic devices, optical techniques are generally restricted by the extremely weak magneto-optical activity and diffraction limit. Highly sophisticated, expensive electron microscopy and scanning probe methods thus have come to the forefront[7,8,9,10,11,12]. Here, we show that perfect optical absorption (POA) dramatically improves the performance and functionality of MOKE microscopy. For 1-nm-thin Co film, we demonstrate a Kerr amplitude as large as 20° and magnetic domain imaging visibility of 0.47. Especially, POA-enhanced MOKE microscopy enables real-time detection and statistical analysis of sub-wavelength magnetic domain reversals. Furthermore, we exploit enhanced magneto-optic birefringence and demonstrate analyser-free MOKE microscopy. The POA technique is promising for optical investigations and applications of nanomagnetic systems.

## Main

At the heart of our novel high-performance MOKE microscopy is POA, which is realised by two thin $SiO_2$ spacer layers and a bottom Al mirror, as illustrated in the right panel of Fig. 1a[13,14]. The bottom phase-matching $SiO_2$ layer enables destructive interference, to suppress the non-MO reflection amplitude $r_{xx}$. Depending on the thickness of the phase-matching layer, lights reflected from the magnetic layer and the mirror layer interferes. The suppression in $r_{xx}$ appears in a broad spectral range by the phase-compensation of the top $SiO_2$ layer. In addition, a highly confined electric field in the magnetic medium enhances the MO reflection amplitude $r_{xy}$. For the incident light linearly polarised along the x-direction, the Kerr rotation $\theta_K$ and ellipticity $\varepsilon_K$ are related through the Kerr amplitude $|\theta_K + i\epsilon_K| = |\tan^{-1}(r_{xy}/r_{xx})|$. In our multi-layer structure, POA significantly enhances the Kerr amplitude, up to two orders of magnitude compared with conventional measurements. Such an extreme enhancement of the Kerr amplitude facilitates highly sensitive detection and imaging of magnetised domains (Fig. 1a, insets). By optimising the reflectivity of the bottom mirror and the thicknesses of the phase-matching and phase-compensation layers, the POA condition can be achieved for a wide range of the magnetic layer's complex refractive indices and light wavelengths (Supplementary Information S1).

To verify the expected advantages of POA, we calculated the non-MO and MO reflections of a Co 1-nm / Pt 5-nm film on a $SiO_2$ substrate, and those for the same film embedded in a POA multilayer (Figs. 1b and 1c)[15]. We assume that the magnetisation of the Co/Pt medium was saturated along the $+z$ direction[16] and a 660-nm-wavelength $x$-polarised planewave was normally incident from the top. The thicknesses of the top and bottom $SiO_2$ layers are set to 265 nm and 113 nm, respectively, to support POA at the target wavelength (Supplementary Information S2). The bare Co/Pt film on the $SiO_2$ substrate reflects ~45.8% of the incident light back to the air and transmits ~37.7% to the substrate. This results in a weak non-MO field (~0.316) at the Co/Pt layer, which generates a weak MO field (~$1.14\times10^{-3}$). On the other hand, the POA multilayer reflects only ~$7.98\times10^{-4}$%, which results in a strong non-MO field (~0.879) at the Co/Pt layer with a highly enhanced MO field (~$7.37\times10^{-3}$). As a result, the theoretical calculation predicts that the MOKE amplitude of the POA Co/Pt film can be ~203 times larger than that of the bare Co/Pt film.

We first experimentally demonstrate POA-enhanced polar-MOKE for the Co/Pt film (Fig. 2a). Identical magnetic films were simultaneously deposited on a $SiO_2$ substrate and a $SiO_2$/Al/$SiO_2$ substrate, for POA. A 2-nm-thick $AlO_x$ layer on the Co/Pt film was additionally deposited in situ, to prevent oxidation. We achieved high-quality POA, reaching 99.94%. Such high-quality POA suppressed the non-MO reflection amplitude $|r_{xx}|$ down to 0.023, which was ~20 times smaller than for the bare film (Fig. 2b). As predicted theoretically, our platform also enhanced the MO reflection amplitude $|r_{xy}|$ by 3.5 times (on average, over the examined range of wavelengths) larger than for the bare film (Fig. 2e). As a result, the Kerr amplitude in terms of $\tan^{-1}(r_{xy}/r_{xx})$ reached 20.1° at the POA wavelength (660 nm), which was ~66 times larger than that of the bare Co/Pt film, and was >10° over a wide spectral range, from 648 nm to 674 nm.

The enhancement and isolation of the MOKE signal by POA are universal, regardless of the type of magnetic medium. We designed and fabricated a POA multilayer structure for a Pt/Co/Pt/Ta film (Fig. 2c), which exhibits different magnetic domain reversal dynamics from the Co/Pt film. Here, although imperfect fabrication yielded relatively low-quality POA of 97.14%, the mechanism for isolation and simultaneous enhancement of the MOKE signal was still effective; the non-MO reflectivity amplitude was suppressed by 3.98 times, while the MO reflectivity amplitude was enhanced by 2.93 times (Figs. 2f and 2g, respectively). Even without high-quality POA, the Kerr amplitude reached ~0.98°, which was ~11.2 times higher than that for the bare Pt/Co/Pt/Ta film (Fig. 2h).

The POA platform significantly improves the performance of the MOKE microscopy. As illustrated in Fig. 3a, the magnetic domains with opposite vertical magnetisations (+M, −M) exhibit different MOKE intensities ($I_{MOKE}$) through an analyser of angle ($\delta$) (Supplementary Information S3). The relative phase difference between the non-MO and MO reflection can be compensated by installing a waveplate in front of the analyser[13]. The non-MO reflection determines the reference value of $I_{MOKE}$ (indicated by the yellow line), and the MO reflection increases or decreases $I_{MOKE}$ (indicated by the red and blue lines), depending on the direction of magnetisation. High-visibility MOKE measurements demand a wide intensity detection range ($\Delta I_{MOKE}$), ideally approaching the full-dynamic range of the employed detector. Typical high-performance analysers/waveplates support high visibility for a few mrad MOKE signal, within a very narrow working range of the analyser angle. On the other hand, POA enables high-visibility MOKE measurements over a wide range of the analyser angle, even employing a low-extinction analyser.

Figures 3b and 3c show the calculated visibilities of MOKE measurements for the bare and POA films, depending on the extinction efficiency and the analyser angle (Supplementary Information S4). For the bare Co/Pt film, although a high-extinction analyser ($\eta = 10^5$) was used, the visibility was only ~0.047. On the other hand, the POA Co/Pt film, with the high-extinction analyser, showed high MOKE visibility of 0.72 at the peak and values >0.10, over a wide range of the analyser angle, from 1.2° to 79°. Besides, the use of a low-extinction analyser ($\eta = 10^2$) still supports high visibility of 0.46 at the peak and values >0.10, for the analyser angle in the 2.5°–78° range, which still exceeds the values obtained using the high-extinction analyser for the bare Co/Pt film. Despite its imperfect performance, the POA Pt/Co/Pt/Ta film's visibility reaches 0.23 and 0.042, for the high- and low-extinction analysers, respectively.

We demonstrated high-visibility MOKE imaging of two representative magnetisation reversal processes in ferromagnetic media: domain nucleation (annealed Co/Pt film, Fig. 3d) and domain wall propagation (non-annealed Pt/Co/Pt/Ta film, Fig. 3e)[17,18]. In the experiment, all of the magnetic domains of the 1-nm-thin Co medium were magnetised initially in the +z direction, and then an external magnetic field (320 G for Co/Pt, 100 G for Pt/Co/Pt/Ta) in the -z direction was applied to trigger magnetisation reversal. For the Co/Pt film, POA MOKE imaging under a 660±5 nm light-emitting diode illumination resolves the magnetic domains with opposite vertical magnetisations, with high visibility of 0.47 on average, over the field of view. The angle and extinction ratio of the employed analyser were 10° and ~$10^3$, respectively. The combination of the high visibility and the large analyser angle increases the accuracy of solving the quadratic equation for extracting vertical magnetisation from the MOKE intensity data (Supplementary Information S3). We were able to determine not only the direction but also the magnitude of the net vertical magnetisation of the magnetic domains within the resolution area of the objective. In the POA Pt/Co/Pt/Ta film, the domain wall propagation, showing irregular development owing to the Barkhausen jumps, was clearly measurable, with high visibility of ~0.14. Note that, under the same measurement conditions, the visibility of the bare Co/Pt (Pt/Co/Pt/Ta) film was only ~0.025 (~$4.6\times10^{-4}$) (Supplementary Information S5).

The most innovative result of this study is the optical measurement of the statistics of the Barkhausen jumps[19,20] in the sub-wavelength regime. As illustrated in Figs. 4a and 4b, the intensity of the reflected MOKE signal changes depending on the area ratio of the opposite magnetisation domains inside the detection area (indicated by the red dotted circle). In real-time measurements, the Barkhausen jumps are observed in the form of abrupt stepwise changes of the MOKE intensity (Figs. 4c and 4d). We converted the stepwise intensity change ($\delta I$) to the area of the reversed magnetic domain ($\delta A$) based on the quadratic equation (Supplementary Information S3). It is however notable that, when employing a large analyser angle (e.g., 40° used for the results in Fig. 4), the MOKE intensity changes almost linearly with the reversed domain area, as $\delta I / \Delta I_{\text{MOKE}} \cong \delta A / \pi R^2$. Here, $\Delta I_{\text{MOKE}}$ is the MOKE intensity change when all magnetic domains in the detection area of $\pi R^2$ are reversed completely. In experiments, a 660-nm-wavelength laser illuminated the samples, and a pinhole cropped the signal from an area with the diameter (2$R$) of 1.33 μm. The incident laser power was ~110 μW, sufficiently weak to avoid thermal effects but sufficiently strong to use almost the full dynamic range of the employed detector (Supplementary Information S6). The detection noise fluctuation, which is the resolution limit of the MOKE intensity change measurement, determine the minimal size of measurable domains ($\sigma$). Owing to its high visibility, the POA MOKE measurement indeed enabled to resolve the magnetisation reversal of domains with areas as small as ~1.11×10$^{-3}$ μm$^2$ (~4.57×10$^{-3}$ μm$^2$) in the Co/Pt (Pt/Co/Pt/Ta film), beyond the optical diffraction limit.

We temporally recorded the Barkhausen jump events depending on the area of the reversed magnetic domain (Figs. 4e and 4f). We plotted 7379 (2682) events for the POA Co/Pt (Pt/Co/Pt/Ta) film obtained from 50 real-time measurements under the same experimental condition. The measurement of the size-dependent dynamics of magnetisation reversal reveals the distinction between the domain nucleation and domain wall propagation processes. As shown in Fig. 4e, the dynamics of domain nucleation exhibits temporal exponential decay, with the reaction rate $k$. The Arrhenius equation gives $k \propto \exp(-E_a/k_B T) = \exp\left((K_u - \vec{M} \cdot \vec{B})V/k_B T\right)$, where $K_u$, $\vec{M} \cdot \vec{B}$, and $V$ are the perpendicular magnetic anisotropy constant, the Zeeman energy density, and the volume of the magnetic domain, respectively[14]. On the other hand, in the domain wall propagation, interactions between nearest-neighbour domains dominate; thus, the magnetisation reversal is independent on the domain size (Fig. 4f). It is well known that the Barkhausen jumps satisfy power-law scaling of domain sizes[21]. In this work, we for the first time verify the power-law scalability down to the 10-nm-scale, which has not been demonstrated yet using conventional optical techniques (Figs. 4g and 4h). The critical exponents for the examined Co/Pt and Pt/Co/Pt/Ta films were −1.91 ± 0.10 and −1.83 ± 0.13, respectively. We expect that using high-performance photodetectors with a large dynamic range and GHz-level bandwidth will allow further investigations into the size-dependent dynamics of magnetic domain reversal, on the atomic scale and with the nanosecond-scale speed.

The isolation of MO reflection enables analyser-free MOKE microscopy. The MO birefringence causes the differential reflectance of left- and right-hand circularly polarised light[22]. In the Co medium in this study, the domain magnetised in the +z direction reflected left-hand circularly polarised light more than right-hand circularly polarised light, and vice versa. As shown in Fig. 5a, the MO birefringent reflection emerges from the non-MO reflection near the POA condition. The ratio of the reflectance of the right- and left-handed circularly polarised light was up to ~1.9, corresponding to the visibility of ~0.30 (Fig. 5b). Such a high circular birefringence enables straightforward but powerful MOKE microscopy without any analyser or waveplate. The bright-field microscopy, using only circularly polarised illumination, excellently resolved the regions magnetised in the +z and −z directions (Fig. 5c). Analyser-free MOKE detection can also facilitate the accessibility of MO devices.

In summary, POA provides a simple, promising way for significantly expanding the conventional limits of MOKE microscopy. High-visibility MOKE microscopy based on the remarkably enhanced Kerr amplitude enables real-time measurement and statistical analysis of nanoscale magnetic domain reversal beyond the optical diffraction limit. In the future, we expect to use the POA MOKE microscopy approach for investigating state-of-the-art topics in the field of magnetism and spintronics, such as skyrmions[23], magnons[24], and magnetic solitons[25,26]. We also believe that the POA technique will significantly facilitate utilisation of optically functional nanoscale-thin films, from semiconductor quantum wells[27] to two-dimensional materials[28,29], for novel photonic and electro-/magneto-optic devices and systems[30].

## Methods

### Fabrication

The 100-nm-thick Al mirror layer for POA was deposited by the conventional electron beam evaporation. The phase-matching and phase-compensation $SiO_2$ layers were deposited by the radio-frequency sputtering process. We employed physical vapour deposition by sputtering, to form the Co/Pt and Pt/Co/Pt/Ta layers. The magnetic layers of the bare and POA samples were grown at the same time. To support the reverse domain nucleation behaviour, we performed post-annealing treatment on the Co/Pt layer at 350 °C.

### Measurement

The fabricated samples were mounted on a precise, non-magnetic XYZ translation stage at room temperature. A permanent neodymium magnet with a magnetic field strength of ~5000 G at the surface was employed to apply an external magnetic field along the direction normal to the magnetic layer. A motorised translation stage moved the magnet and controlled the strength of the magnetic field applied to the target sample, up to 5000 G. Employing a Hall effect sensor (Hirst Magnetic Instruments GM08), we mapped the direction and strength of the magnetic field depending on the distance relative to the magnet. In spectral measurements, a broadband light-emitting diode ($\lambda$ = 640 nm, $\Delta\lambda$ = ±50 nm) illuminated the sample, and a spectrometer (Princeton Instruments Acton SP2300) and high-performance charge-coupled device (Princeton Instrument PIXIS-100BR) were used. Here, to obtain a planewave-like illumination, we used a ×5 objective lens with a low numerical aperture (NA) of 0.1. In MOKE imaging, a narrowband light-emitting diode (LED) ($\lambda$ = 660 nm, $\Delta\lambda$ = ±5 nm) illuminated the target sample, and a high bit-depth complementary metal-oxide-semiconductor (CMOS) camera (Sony IMX249LLJ-C) was used. For the Barkhausen jump measurements, a 660-nm-wavelength laser diode and a femtowatt photoreceiver (Newport 2151) were used. We used a ×50 long-working-distance objective lens (Mitutoyo, NA 0.42) for MOKE imaging and for the Barkhausen jump measurements.

### Theoretical calculations

The permittivity of the MO medium for theoretical calculations was modelled by a Hermitian tensor, consisting of the diagonal non-magnetic permittivity and off-diagonal magnetic components. The non-magnetic permittivity of the Co/Pt and Pt/Co/Pt/Ta media was obtained from ellipsometry measurements of the bare films. The off-diagonal magnetic components were extracted from the MOKE measurements of the bare films. The permittivity of the bare films was then used to calculate the reflection amplitude ($r_{xx}$, $r_{xy}$) spectra and MOKE spectra, using the anisotropic transfer matrix method[31] (Fig. 2). The permittivity of the deposited $SiO_2$ layer was also obtained from the ellipsometry measurements. We used the permittivity of Al from experimental measurements[32]. The electric field profiles (Figs. 1b and 1c) were calculated using the finite-difference time-domain (FDTD) simulation (Lumerical Solutions, Inc). We used the grid attribute technique of the employed FDTD calculator to deal with the Hermitian permittivity tensor of the magnetic medium.

**Acknowledgements** M.-K.S. acknowledges support by the National Research Foundation of Korea (NRF) (2017R1A2B2009117, 2017R1A4A1015426, and 2014M3C1A3052537). K.-J.K. acknowledges support by NRF (2017R1C1B2009686). B.-G.P. acknowledges support by NRF (2017R1A2A2A05069760). J.S. acknowledges support by NRF (2017R1A2B2005702).

**Author Contributions** D.K., and M.-K.S. designed the concepts and experiments. Y.O., and J.K. fabricated the experimental samples. D.K. performed the spectral measurements, MOKE imaging, confocal MOKE measurement, and analyser-free MOKE microscopy. D.K. performed optical simulations. D.K., K.-J.K., and M.-K.S. wrote the paper. M.-K.S. supervised the project. All authors discussed the results and commented on the manuscript.

**Author Information** The authors declare no competing financial interests. Correspondence and requests for materials should be addressed to M.-K.S (minkyo_seo@kaist.ac.kr).

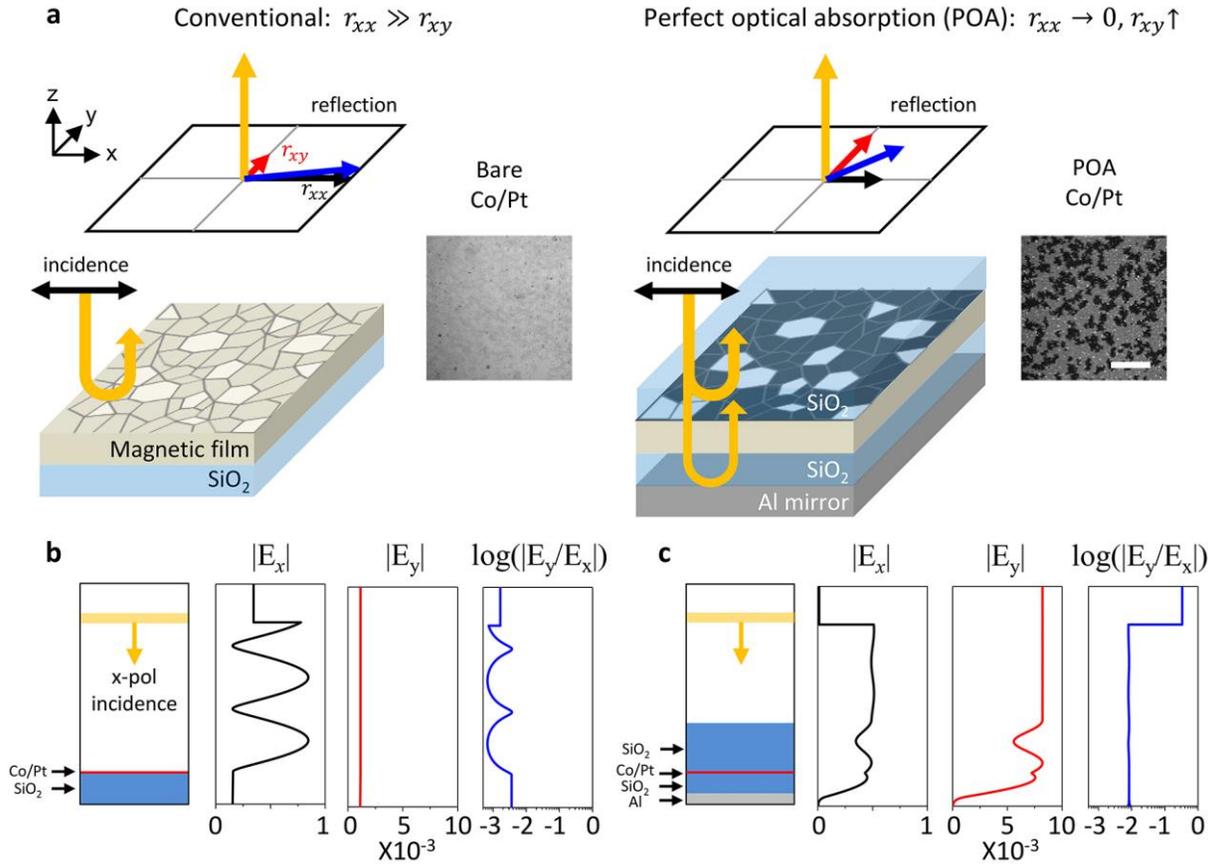

**Figure 1 | POA enhanced MOKE microscopy. a,** Schematic of the conventional and POA-enhanced MOKE microscopy. The cooperation of the bottom Al reflector, phase-matching $SiO_2$ layer, and phase-compensation $SiO_2$ layer presents the magnetic material with POA (right panel). POA not only suppresses the non-MO reflection of the incident field ($E_x$) but also enhances the generation of the MO field ($E_y$). The combination of the suppressed ($r_{xx}$) and enhanced ($r_{xy}$) dramatically improves the performance of the MOKE microscopy. Insets, MOKE microscopy images of the bare and POA Co/Pt films. Scale bar, 50 μm. **b,c,** Simulated electric field amplitude profiles of the bare Co/Pt **(b)** and POA Co/Pt **(c)** films under the normal incidence of the *x*-polarised planewave. The yellow line represents the position of the planewave source. The wavelength of light is 660 nm. The MO Kerr amplitude, the ratio between $|E_x|$ and $|E_y|$, is plotted on the log scale.

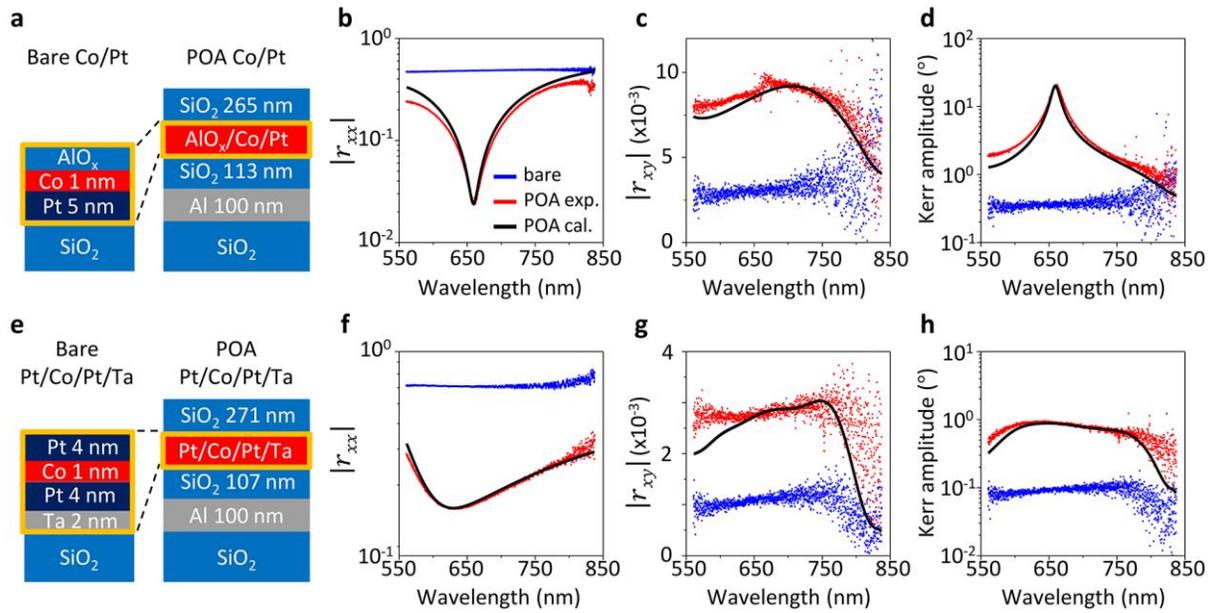

**Figure 2 | Experimental demonstration of POA-enhanced MOKE. a,** Schematics of the bare and POA Co/Pt layers. **b-d,** Measured spectra of the non-MO reflection amplitude $|r_{xx}|$, MO reflection amplitude $|r_{xy}|$, and Kerr amplitude from the bare and POA Co/Pt layers. **e,** Schematics of the bare and POA Pt/Co/Pt/Ta layers. **f-h,** Measured spectra of the non-MO reflection amplitude $|r_{xx}|$, MO reflection amplitude $|r_{xy}|$, and Kerr amplitude from bare and POA Pt/Co/Pt/Ta layers. We also calculated the non-MO and MO reflection amplitudes and the Kerr amplitude of the POA Co/Pt and POA Pt/Co/Pt/Ta films using the transfer matrix method (black solid lines).

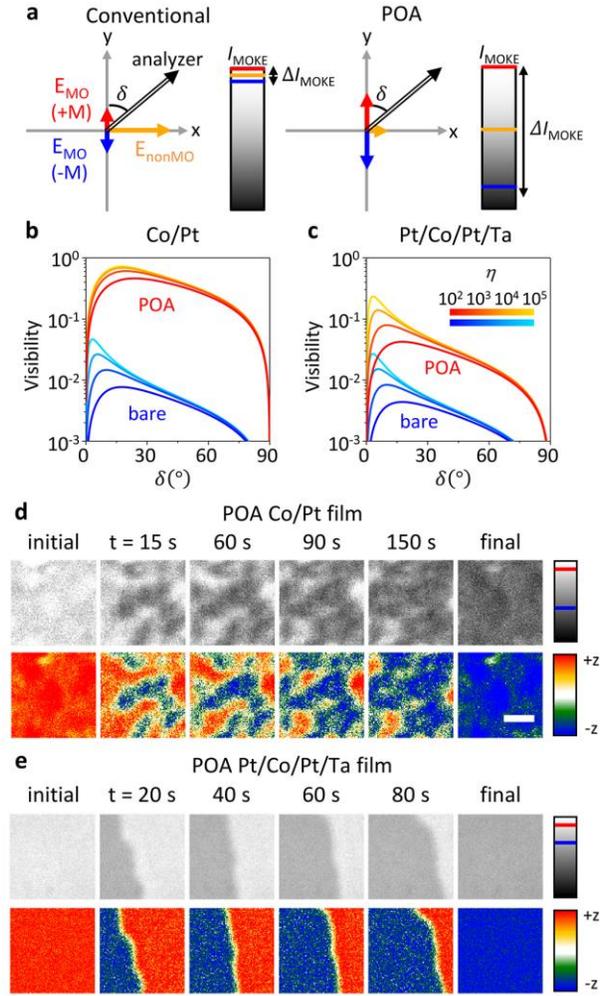

**Figure 3 | High-visibility MOKE microscopy. a,** Schematic of the realisation of high-visibility MOKE microscopy by POA. The yellow arrow represents the electric field of the non-MO reflection. The red and blue arrows indicate the electric field of the MO reflection from the magnetic media of magnetisation +M and –M, respectively. Here, +M (–M) is the magnetisation of the fully magnetised medium in the +z (–z) direction. The MOKE intensity ($I_{MOKE}$) is measured using an analyser with the angle $\delta$ (double-lined arrow). In the colour bar, the non-MO reflection yields the reference value of $I_{MOKE}$ (the yellow line) and the MO reflections of +M and –M determine the maximum and minimum of $I_{MOKE}$ (the red and blue lines). $\Delta I_{MOKE}$, the difference between $I_{MOKE}$(+M) and $I_{MOKE}$(–M), determines the available dynamic range of the employed detector and the visibility of the MOKE measurement. **b,c,** Calculated visibility of MOKE microscopy ($V_{MOKE}$) depending on the extinction ratio ($\eta$) and angle ($\delta$) of the analyser (Supplementary Information S4). **d,e,** Images of the measured MOKE intensity ($I_{MOKE}$) and extracted magnetisation ($m$) of magnetic domain reversal, for the POA Co/Pt **(d)** and POA Pt/Co/Pt/Ta **(e)** films. The red and blue lines in the colour bar of $I_{MOKE}$ indicate the levels of $I_{MOKE}$(+M) and $I_{MOKE}$(–M). Scale bar, 3 $\mu$m.

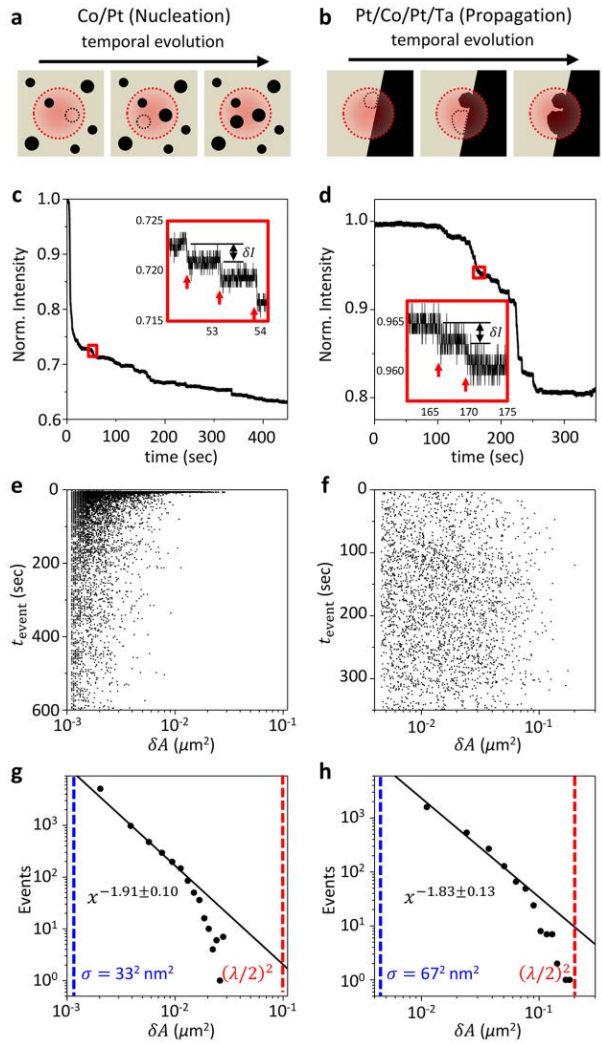

**Figure 4 | Barkhausen jumps beyond the diffraction limit. a,b,** Barkhausen jumps, the magnetisation reversals of individual coherent magnetic domains, cause abrupt changes of the MOKE intensity. The dotted red circle indicates the area of detection. The **(a)** Co/Pt and **(b)** Pt/Co/Pt/Ta films exhibit reverse domain nucleation and domain wall propagation, respectively. **c,d,** Real-time measurements of magnetisation reversals of the **(c)** POA Co/Pt and **(d)** POA Pt/Co/Pt/Ta films. As shown in the magnified plots (insets), the Barkhausen jumps far below the wavelength scale (red arrows) are successfully resolved in terms of the MOKE intensity change ($\delta I$). **e,f,** Temporal statistics of the Barkhausen jumps depending on the size ($\delta A$) in the POA Co/Pt **(e)** and POA Pt/Co/Pt/Ta **(f)** films. **g,h,** Distribution of the Barkhausen size ($\delta A$) in POA Co/Pt **(g)** and POA Pt/Co/Pt/Ta **(h)**. The scale-invariant Barkhausen jump events are fitted by a power law for the domain size (solid line). The red and blue dashed lines indicate the optical diffraction limit and the minimal size of measurable domains ($\sigma$).

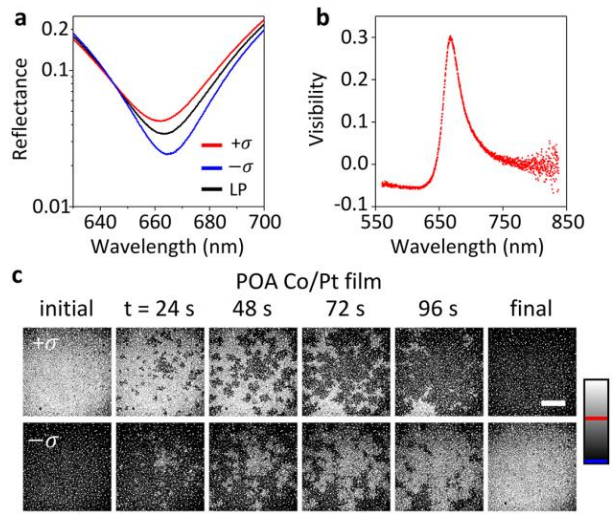

**Figure 5 | Analyser-free MOKE microscopy. a,** Measured reflectance spectra of the POA Co/Pt film under the right- ($-\sigma$) and left-handed ($+\sigma$) circularly polarised incidence and the linearly polarised (LP) incidence. **b,** Visibility spectrum of the circularly birefringent reflectance. **c,** Measured analyser-free MOKE microscope images of magnetisation reversals of the POA Co/Pt film under the left- and right-handed circularly polarised illuminations. The red and blue lines in the colour bar indicate the maximal and minimal values of the measured MOKE intensity, respectively. Scale bar, 50 $\mu$m.